\documentclass{article}
\usepackage{ijcai19}

\usepackage{times}
\usepackage{soul}
\usepackage{color}
\usepackage{url}
\usepackage[hidelinks]{hyperref}
\usepackage[utf8]{inputenc}
\usepackage[small]{caption}
\usepackage{graphicx}
\usepackage{amsmath}
\usepackage{booktabs}
\usepackage{algorithm}
\usepackage{algorithmic}
\usepackage{caption}
\usepackage{subcaption}
\title{Demonstration of PerformanceNet: A Convolutional Neural Network Model for Score-to-Audio Music Generation}

\author{
    Yu-Hua Chen, Bryan Wang, and Yi-Hsuan Yang
    \affiliations
    Research Center for IT Innovation, Academia Sinica, Taiwan \emails
    \{cloud60138, bryanw, yang\}@citi.sinica.edu.tw
}

\begin{document}

\maketitle

\begin{abstract}
  We present in this paper PerformacnceNet, a neural network model we proposed recently to achieve score-to-audio music generation.
  The model learns to convert a music piece from the symbolic domain to the audio domain, 
  assigning performance-level attributes such as changes in velocity automatically to the music and then synthesizing the audio. The model is therefore not just a neural audio synthesizer, but an AI performer that learns to interpret a musical score in its own way.
  The code and sample outputs of the model can be found online at  \url{https://github.com/bwang514/PerformanceNet}.
\end{abstract}

\section{Introduction}


Music is generally considered as organized sounds created by human and is transmitted as audio waveforms. People have designed musical symbols to notate various aspects of music. Accordingly, we can transcribe a sound recording in a handwritten or printed form, facilitating the communication of the ``content'' of the music. However, given the same musical score sheet, different musicians can interpret the music in different ways and use their personal ``styles'' while performing the music. Such performance-level attributes of the music are usually easier to find directly in the audio waveform, not in the symbolic music notation.

Recent years have witnessed a growing interest in building machine models for music generation. However, most existing work focuses on only one of the two main domains of music---symbolic or audio---rather than the two domains at the same time. 
People working on \emph{symbolic-domain music generation}, a.k.a. algorithmic composition, typically focus on generating original musical content such as melody and chords and tend to use off-the-shelf audio synthesizers to play the music they generate (e.g., \cite{midinet,musegan,brunner2018midi,simon18ismir}).  
And, people working on \emph{audio-domain music generation} usually focus on the synthesis part only and aim to generate original sounds of whatever musical content (e.g., \cite{nsynth,gansynth,tfgan}).
There are some things in between that cannot be modeled without considering data from the aforementioned two domains together, such as the performance-level attributes and playing styles.

\begin{figure}[t]
     \centering
     \begin{subfigure}{0.23\textwidth}  
         \includegraphics[scale=0.085]{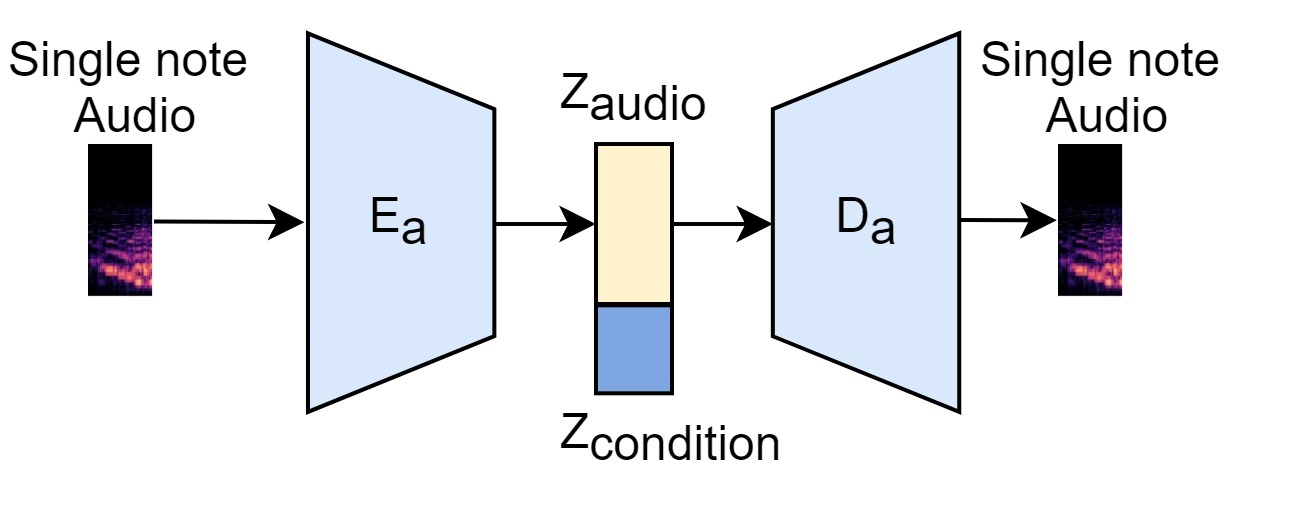}
         \caption{Neural audio synthesizer}
         \hfill
     \end{subfigure}
     \hfill
     \begin{subfigure}[b!]{0.23\textwidth}  
         \includegraphics[scale=0.085]{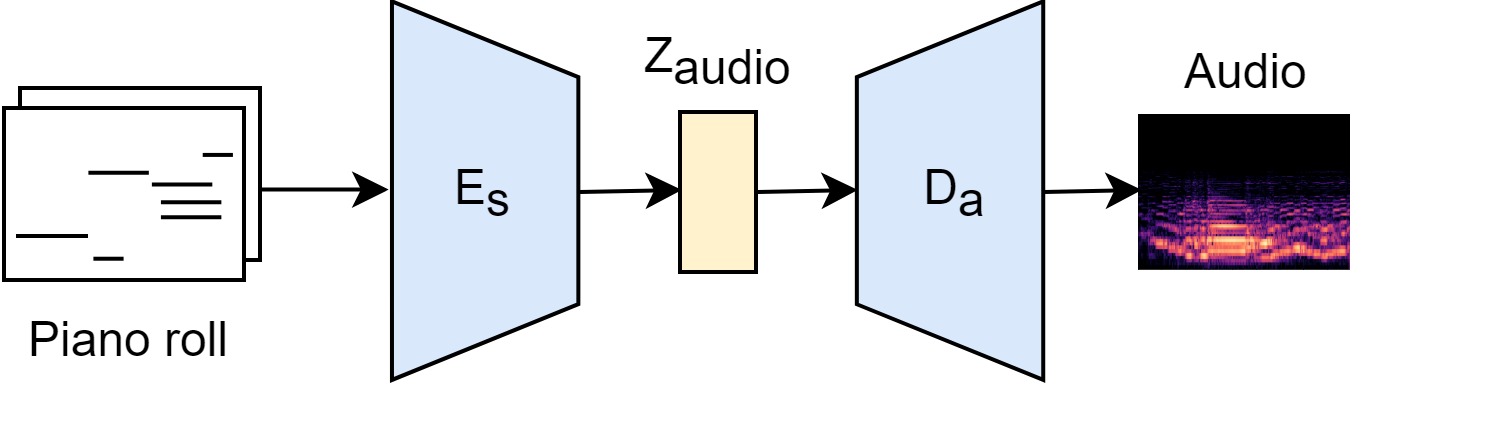}
         \caption{PerformanceNet}
         \hfill
     \end{subfigure}
     \hfill
     \begin{subfigure}{0.5\textwidth}  
         \centering
         \includegraphics[scale=0.1]{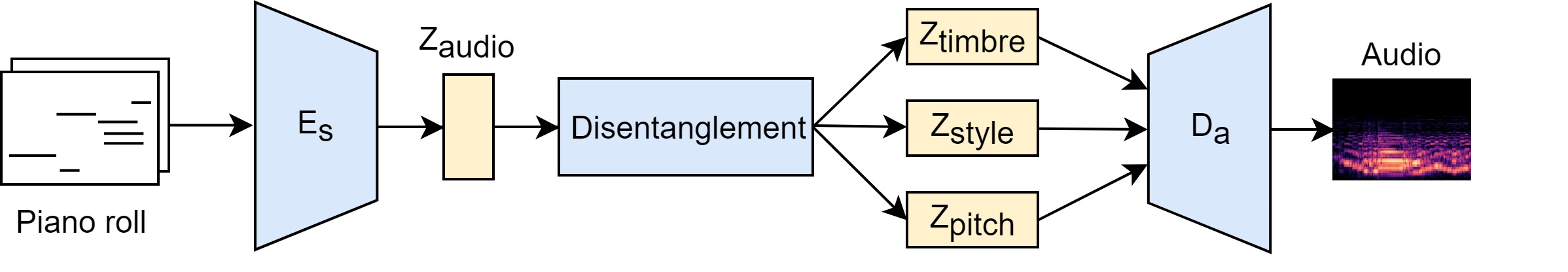}
         \caption{Envisioned extended version of PerformanceNet}
     \end{subfigure}
     \hfill
    \caption{Diagram of encoder/decoder networks for different tasks; E, D, z denote the encoder, decoder and latent code, respectively.}
\label{fig:sys}
\end{figure}

In a prior work, we address this gap by proposing a neural network model, dubbed the ``PerformanceNet,'' that takes symbolic representations of a music piece as input and generates as output a sound recording playing that piece expressively \cite{performancenet}. 
The goal of  PerformanceNet is to predict the performance-level attributes, such as changes in velocity (i.e., dynamics/loudness) and modulations in pitch (e.g., \emph{vibrato}) that a 
human performer may apply while performing the music. 
As shown in Figure \ref{fig:sys}(b), the model also learns to synthesize audio in an end-to-end manner. 
To our knowledge, PerformanceNet, and the work presented independently and  concurrently to our work in \cite{kim19icassp}, represent the first models that learn explicitly 
the score-to-audio mapping of music, for arbitrary instruments.

In this demonstration, we discuss the difference between the \emph{note-level synthesis} task addressed by existing neural audio synthesizers (e.g., \cite{gansynth}) and the \emph{phrase-level synthesis} task addressed by PerformanceNet. 
We also present a graphical user interface with which people can load and edit a musical score and then ask our PerformanceNet to perform it expressively using different instruments.

\begin{figure}[t]
\centering
\includegraphics[scale=0.5]{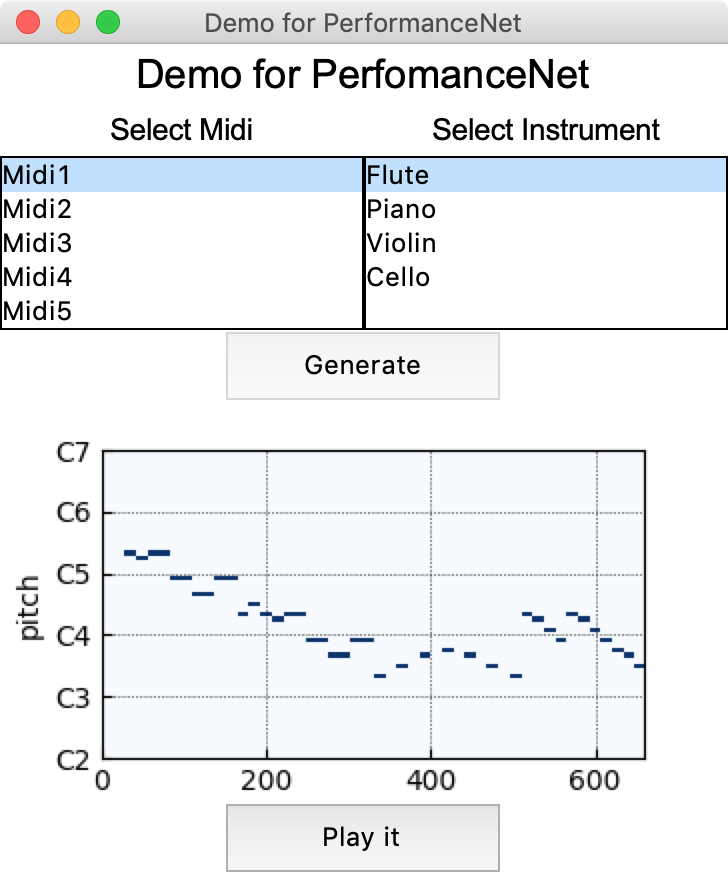}

\caption{A simple GUI for demonstrating PerformanceNet.}
\label{fig:demo}
\end{figure}

\section{Note-level \& Phrase-level Audio Generation}



Most existing neural audio synthesizers
employ a neural network model to learn to generate high-quality musical sounds. The model is usually trained with \textbf{audio recordings of isolated musical notes} (e.g., \texttt{C4} and \texttt{C5}) from different instruments.
Therefore, we refer to them as performing \textbf{note-level audio generation} (or synthesis).
When an encoder/decoder architecture is used, as the case in \cite{nsynth} (see Figure \ref{fig:sys}(a)), the model learns in the latent space the embeddings of the musical timbre of different instruments (marked as $z_\text{audio}$ in Figure \ref{fig:sys}(a)). It is therefore possible to sample from the latent space to create sounds of new instruments, or to interpolate the sounds of existing instruments. To control the pitch, both \cite{nsynth} and \cite{gansynth} concatenate with the latent code a one-hot vector (marked as $z_\text{condition}$ in Figure \ref{fig:sys}(a)) representing the pitch of the sound to be generated. The pitch vector is one-hot (i.e., only one element of the vector takes the value one and the rest are zero), as the models synthesize audio one note at a time. The sounds generated can be realistic and expressive, since the model is trained with real-world audio recordings.

An AI performer, on the other hand, learns to  convert a music piece from the symbolic domain to the audio domain,  assigning performance-level attributes such as changes in velocity automatically to the music and then synthesizing the audio. 
The model is trained with \textbf{pairs of the symbolic representation and audio recordings of musical phrases} comprising of multiple notes.
The input representation used by PerformanceNet, for example, is a 
symbolic representation 
called the \emph{pianoroll} \cite{pypianoroll}, a binary, scoresheet-like matrix representing the presence of notes over different time steps for a single instrument. We can extend it to a tensor, i.e., multitrack pianoroll, to represent the score of multiple instruments. The output representation used by PerformanceNet is the (magnitude) spectrogram of the corresponding audio recording, so what PerformanceNet learns is actually a matrix-to-matrix mapping. When an encoder/decoder architecture is used, as the case in PerformanceNet (illustrated in Figure \ref{fig:sys}(b)), the latent code contains not only timbre but also style and pitch information. Therefore, disentanglement techniques \cite{hung18arxiv} may be needed to disentangle these elements, as illustrated in Figure \ref{fig:sys}(c).

We can now see that a core task of an AI performer is therefore \textbf{score-informed phrase-level audio generation}.
Unlike the case of note-level generation, here we need to learn how to connect different notes while playing (e.g., using playing techniques such as \emph{slide}, \emph{hammer-on} and \emph{pull-off} as the case in guitar music \cite{chen15ismir,tent}), and  to play the same pitch differently depending on the position of that note in a phrase (e.g., whether it is at the downbeat) \cite{li15ismir}. Moreover, an AI performer holds the potential to learn better the phrase-level attributes 
of music, and accordingly the playing style of different musicians \cite{shih17dafx}. This might be done, for example,  by conditioning the PerformanceNet with an one-hot vector indicating the musician who played that phrase.

\section{Model Architecture of PerformanceNet}


The PerformanceNet consists of two subnets.
The first subnet, the  ContourNet, uses a convolutional encoder/decoder architecture to roughly convert the pianoroll to the spectrogram. 
The second subnet, the TextureNet, further improves the result of the ContourNet by refining the details of the partials of each note in the spectra with convolutional layers of a multi-band residual design.\footnote{We found that TextureNet's refinement is two-fold. Firstly, it sharpens the blurred frequency bins close to the fundamental frequency, which contributes to better reconstructed audio quality as pointed out in \cite{timbretron}. Secondly, overtones with higher frequencies, which contribute to the perception of realistic timbre, are gradually added to the spectrogram by multi-band residual blocks, demonstrating the coarse-to-fine rendering process. We show figures demonstrating these in our project website.} 
The job of the ContourNet is akin to performing \emph{domain translation} \cite{gatys16cvpr} (between the symbolic and audio domains of music), whereas the TextureNet is doing \emph{super resolution} \cite{ledig17cvpr}. Please see \cite{performancenet} for more technical details.

\section{Demo System}
For the purpose of demonstration, we build a graphical user interface for PerformanceNet, as depicted in Figure \ref{fig:demo}. Users can select a MIDI file or upload one. After the score is given, users can choose the instrument to play the piece. 
The audio can be generated on-the-fly by our model.

\bibliographystyle{named}
\bibliography{ijcai19}

\end{document}